\documentstyle[12pt]{article}     
\textwidth  6.22in  \textheight  8.5in  \date{(Date:  28.  07.  2009)}
\begin{document}
\title{\Large   On   Reality    of   Tachyonic   de   Broglie   Waves}
\author{Rajat K. Pradhan\thanks{Present  address: V. Dev College, Jeypore
    - 764   001,    India}   \thanks{Corresponding   author;   E-mail:
    rajat@iopb.res.in}~ and Lambodar P. Singh} \maketitle {\it Department of
  Physics, Utkal University, Bhubaneswar - 751 004, India.}

\vspace*{0.2in}

\hspace*{2.2in} {\bf Abstract}
\vspace*{0.1in}

We investigate  the tachyonic  nature of the  de Broglie  matter waves
associated with a free quantum object to show that granting reality to
them would  lend completeness to  the quantum description  of reality.
Basing  on  the  robustness  of  the well  known  Einstein-de  Broglie
reciprocal relation between the  phase and the particle velocities, we
extend the  concept of complementarity  to them and thereby  propose a
complementary relation between a bradyon and its corresponding tachyon
(i.e. the associated matter wave)  to endow the tachyons with a degree
of  reality, at least  on par  with the  bradyons, within  the current
framework of  quantum physics and extended relativity.  The duality is
used to argue that because of the observed localised nature of bradyons,
tachyons should always  be pervasive or global in  character and thus,
there can be no  point-like tachyons. A common misconception regarding
the  nonrelativistic  limit of  the  Einstein-de  Broglie relation  is
pointed out and the consequent error of long standing is remedied.

\vspace{0.3in} PACS: 03.65Ca, 03.65Ta, 03.75.-b.\\
\hspace*{0.25in}Key  words:  foundations  of  Quantum  theory,  matter
waves, measurement, tachyons.
\newpage
\noindent
{\bf 1. Introduction}\\

Tachyons  \cite{recami} have  been the  focus of  attention  in recent
times in  three different sectors.  Firstly, the researches  in String
Theory have  led unequivocally to  their presence and  currently there
has been  great deal of interest in  tachyon condensation \cite{asoke}
on   branes   and    tachyonic   inflation   \cite{infl}   in   String
cosmology. Secondly, the tachyons have been investigated as candidates
for     the     dark     matter    in     relativistic     cosmologies
\cite{padman}. Thirdly,  there have  been analyses and  experiments on
superluminal transmission in evanescent photon tunneling as well as in
quantum mechanical barrier  penetration phenomena \cite{reca}. In this
work, we  look for  a possible place  for tachyons within  the current
framework of quantum theory  following de Broglie's original treatment
of wave-particle duality \cite{deBrog}. \\

The traditional description of a free quantum object has been in terms
of a wave  packet composed of an infinite number of  plane waves so as
to grant meaning to the observed localized nature of the object within
the limits  set by  the uncertainty principle.  The duality  between a
particle  and  its corresponding  wave  envisaged  by  de Broglie  was
eschewed  in  favour  of the  description  in  terms  of a  packet  of
probability waves in the interpretation of quantum theory developed by
Born et al., which is the standard quantum theory. \\

The basic problem with a  single de Broglie wave is its non-localized
nature  which  gives an  infinite  extendedness  to  the free  particle
contrary  to  its {\it  observed}  localized  nature. The  uncertainty
principle on the other hand, achieves the localization with a position
uncertainty  but,   at  the  same  time,   introduces  a  simultaneous
uncertainty in the momentum. A second problem with the de Broglie wave
is the fact that its velocity, which is the wave or phase velocity, is
not equal  to the particle  velocity and for material  particles which
always move with subluminal speeds (bradyons), it exceeds the speed of
light. Therefore,  these superluminal  matter  waves are  undoubtedly
tachyonic in  nature. Some  experiments with tunneling  phenomena have
been performed to ascertain  actual superluminal transmission across a
potential barrier \cite{expt}. \\

In   traditional  relativity,   superluminal   motion  is   considered
unphysical because  of the  restrictive second Einstein  postulate and
since tachyons have not  been observed experimentally, the wave packet
description is generally accepted to  be the only valid description of
a free quantum  object.  However, it has got its  own problems also as
delineated below:\\
 
$\bullet$~  {\bf  spreading:}  As  is  well  known,  the  wave  packet
inevitably  spreads  in  space  as  the  particle  moves  due  to  the
nonlinearity  of the  energy-momentum relation  and the  harmonic time
dependence  of  the  wave  function.  The  description  becomes  quite
inadequate and  unsatisfactory at  large times.  The  packet disperses
with time occupying ever larger spatial domains.\\

$\bullet$ ~{\bf Interpretation:} As per the Copenhagen interpretation,
these  waves are  probability waves  and we  can only  talk  about the
probability of finding the particle at  a point in space. But, this is
not accepted  by the other proposed interpretations  of quantum theory
like  the de  Broglie-Bohm pilot  wave  interpretation\cite{bohm}.  In
fact, both the path integral approach of Feynman \cite{bohm1}, and the
many worlds/minds interpretation \cite{qm}, envisage the motion of the
particle as  a sum over  all possible particle paths  with appropriate
amplitudes,  not necessarily  confined to  the narrow  tubelike region
traversed by the linear movement of the centre of the wave packet from
one point  to the other as  expected from Ehrnfest's  theorem for wave
packet dynamics. Obviously, Feynman  grants more conceptual reality to
the particle itself compared to the wave packet, since the sum is over
particle  paths and  not over  wavepacket traversal  tubes  during the
motion.\\

$\bullet$~ {\bf Irrationality} The {\it bradyonic} wave packet is made
up  of superposition  of individual  {\it tachyonic}  waves  which are
themselves unphysical  from the viewpoint of relativity.   We may ask:
How  can we  get a  description of  a physical  bradyonic  particle by
superposing  `unphysical' tachyonic  solutions?  It  is  certainly not
logically satisfying  that we label the component  waves as unphysical
and then call  their superposition as physical, for  in that case, all
physicality may be said to  be rooted in utter unphysicality, which is
neither acceptable nor tenable. \\

$\bullet$~  {\bf Extrapolation}  Before  measurement (or  any kind  of
interaction  involving the  object), we  have  got to  assign equal  a
priori probability of its being anywhere in space.  It is important to
remember that ``Our common sense  notion of a point particle is always
a post-measurement  notion."  Because we cannot  say anything definite
about  the pre-measurement  state of  a particle,  we cannot  also say
anything  about the  pre-measurement position  of the  particle, since
position  and  momentum  are   the  essential  observables  for  state
characterization. \\

This is as simple as it sounds and follows from common-sense knowledge
of quantum mechanical principles, since before a measurement we cannot
even know whether a quantum object  is a point object or not. The very
knowledge of the  existence of an object presupposes  a measurement of
its  state  or position.  We  cannot  imagine  an object  without  the
category of space or spatial location associated with it.  Further, in
view of the postulate of quantum measurement in regard to the collapse
of  the  wave  function,  we  may  very well  argue  that  it  is  the
measurement   (or  interaction)   that  brings   about   the  familiar
individuated,  localised   particulate  existence  from   a  pervasive
position wave function with constant amplitude everywhere.\\

Thus,  we have  to endow  the individual  de Broglie  waves  with some
degree of reality,  at least on par with  their superposition i.e. the
wave packet.   This leads us to  a further investigation  of the issue
with a view to clearly  bringing out the essential pervasive character
of  a free  quantum object,  which we  have all  along been  trying to
circumvent   by  resorting   to  a   confined,   pointlike,  localized
description with the help of various adhoc assumptions in the standard
quantum theory. We may ask,`` when in the Feynman picture, we allow for
all possible paths  during the motion, why cannot  we allow, a priori,
all possible  positions to the quantum  object at rest?"   In fact, if
this is  done, it would  lead to a  description quite in  keeping with
what  the  de  Broglie  relation  would  imply  for  the  particle  at
rest($\lambda\rightarrow\infty$). Of  course, Feynman's interpretation
of the Dirac factor $exp  (iS/\hbar)$ as the probability amplitude for
the path was to recover  the Born probabilistic interpretation for the
acceptability  of   his  formulation.   But  here,  in   our  proposed
interpretation,  `{\it the particle  actually follows  all  possible paths
unless it is subjected to a measurement to determine its transit route'}. \\

The paper  is organised  as follows:  In Section -  2, we  discuss the
concept of  the free  quantum object  as used in  this paper  and then
derive  the Einstein-de  Broglie relation  between the  phase  and the
particle velocities. In Section - 3, we look at the issue of spreading
of the free  particle wave packet taking the  relativistic formula for
energy and study the implications. A common misconception in textbooks
as well as  in some recent works regarding  the non-relativistic phase
velocity being half  the group velocity are clarified  in the light of
this  derivation in  Section -  4 and  the uncontradictability  of the
Einstein-de  Broglie  relation   is  established.   This  relation  is
rederived  from a  different perspective  in Section  - 5,  wherein we
bring  out a reinterpretation  of the  relation.  In  Section -  6, we
propose  an  extension and  a  generalization  of the  complementarity
principle to  bradyon-tachyon duality/complementarity on  the basis of
this re-interpretation.  Finally,  in Section - 7, we  conclude with a
discussion on the new  interpretation of quantum theory presented here
and point out some of its shortcomings and also its advantages.\\

\noindent
{\bf 2. Phase velocity and particle velocity} \\

Consider a  free quantum object  at rest characterized by  rest energy
$E_0$. The  Schr\"{o}dinger equation $ H  \psi = i d\psi  / dt$ yields
$\psi(t)=$\(A\) $exp(-i\omega_0 t) $ where, $\omega_0= E_0 / \hbar$ is
the   frequency  of   vibration  and   $H=E_0$  is   the  Hamiltonian.
Interestingly enough,  since the Hamiltonian  here is having  no space
dependence,  the  quantum  object  can  be  said  to  be  either  {\it
  independent of space or to be equally pervading all space}. If it is
completely independent of  space, then there is no  possibility of our
ever making any contact with  it except through the time dimension, in
which it is seen to be a harmonic vibration.  The other alternative is
to interpret it  as a cosmic vibration pervading  all space with equal
amplitude  for existence  at all  points.  This  has been  used  by de
Broglie  \cite{debro}  to arrive  at  his  famous relation  expressing
wave-particle duality.\\

For an  observer moving with a  velocity ${\it v}$  along the negative
x-direction,  this cosmic  vibration will  appear to  have  a velocity
${\it  v}$ in the  positive x-direction.  Lorentz transforming  to the
moving frame and on employing
\begin{equation}
t^\prime = \gamma(t-vx/c^2)
\end{equation}
 with $\gamma =  (1-\beta^2)^{1/2}$ and $\beta= v/c$, we  have for the
 wave function:
\begin{eqnarray}
 \psi(t^\prime)=\psi[\gamma(t-xv/c^2)]    =    A~    exp~[-i\omega_{0}
   t^\prime] = A~ exp~[-i\omega_{0}\{\gamma(t-xv/c^2)\}]
\end{eqnarray}
This wave function should have the generic form
\begin{equation}
 \psi(x,t) = A~ exp~[-i\omega_{0}\{\gamma(t-x/v_w)\}]
\end{equation}
 whence  we  get  $v_w=  c^2  /v$  for  the  wave  velocity  or  phase
 velocity.\\
 
This gives  us the Einstein-de  Broglie relation between  the particle
velocity $v$ and the wave velocity $v_{w}$ as :
\begin{equation}
v.v_w =c^2
\end{equation}\\

 Thus, the changed  frequency of the traveling wave  train is given by
 $\omega=\gamma \omega_{0}$  and the changed  energy of the  object is
 $E=\gamma E_{0}=\gamma \hbar \omega_{0}$. The wavelength  is
 $\lambda_{dB}=v_{w} / \nu=c^2  / {v \nu}  $ and on using  the general
 formula $p=vE  / c^2$  for momentum,  we  get the relation $\lambda_{dB}=h/p$.
As  is well known,  the de  Broglie relation  expressing wave-particle
duality had  an undoubted interpretation via the  association of `{\it
  some  kind of  a matter  wave}' with  a matter  particle in  the old
quantum theory before the emergence of the new quantum theory with the
standard  Born probabilistic  interpretation of  the  Schrodinger wave
function.  But, it  was  stripped  of all  its  significance when  the
probability interpretation gained currency.\\
 
The reasons for this are twofold:\\

$\bullet$~ First, It represented a tachyonic wave (associated with the
bradyonic  particle)  which seemed  not  only  to  move ahead  of  the
particle leaving  the latter behind  but also was unphysical  from the
point of view of the second postulate of relativity.\\

 We wish to clarify regarding this objection that {\it the matter wave
   being pervasive along the direction of motion is everywhere present
   and thus cannot ``leave" the particle ``behind" anywhere, for there
   is no position  where the wave is not}. So, we  see that because of
 its tachyonic  character the wave can, without  any contradiction, be
 associated  with  the particle  all  through  its motion.   ``Leaving
 behind" is  a notion applicable  only to bradyons (which  always have
 finite extension) looked at from subluminally moving frames
\footnote{Even for photons(luxons) the concept of ``leaving behind" is
  not applicable  because of the  constancy and the maximality  of the
  speed of  light. For  a photonic(or, infinite--momentum)  frame, the
  entire  line  of  motion  shrinks  to  a  point  because  of  length
  contraction and all entities along the line are seen to be on top of
  each other simultaneously,  unless of course, the line  of motion is
  infinite  in extent  (in which  case, it  may contract  to  a finite
  length).}. \\

$\bullet$ ~  Second, because of the  infinitude of its  extent, it did
not  aid  the  visualization  of  the point  particle  as  a  somewhat
smeared-out existence  in a finite  region of space as  a satisfactory
extension of the concept of  the point particle as required to explain
the various quantum phenomena. \\

Regading  this second point,  we recall  that the  Born interpretation
looks  at  the wave  function  as  representing  the ``probability  of
existence"  at a  space-time point  and not  at ``existence"  as such.
Thus, matter waves become  moving probability waves and physicality or
materiality of the waves is simply washed out with this interpretation
and  what is  retained  is, surprisingly  enough,  only our  classical
notion of the particle as {\it a point object located at some point in
  space at any instant of time}! Having a probability of being located
at a  cerain point  at any  instant of time  precisely means  that the
particle is a point  object. The probability interpretation thus keeps
intact the common-sense notion  of the particulate existence of matter
at  the  fundamental  level,   rendering  our  description  of  matter
classical, but  with the added  factor of probability.   Assuming that
this  widely accepted  interpretation is  correct, we  still  run into
problems with  maintaining the permanence  of the free  quantum object
since  the wave packet  that we  construct for  it by  superposing the
component  probability   waves  inevitably  spreads  in   time  as  it
moves. This  is because of  the nonlinear relation between  energy and
momentum which leads to dispersion. \\

Thus, there  is no contradiction involved in  de Broglie's association
of  a tachyonic  matter wave  with  a bradyonic  particle.  What  this
discussion brings out is that --\\

(a) the quantum  object at rest is actually  a pervasive existence and
when  it  is in  interaction  with  something  else like  a  measuring
instrument, it  appears to be  a finite, localised existence  which we
have all along tried to describe by constructing wave packets.\\

(b) when it is in motion,  it is again a pervasive existence along the
line of motion represented by the de Broglie wave train.\\

In either case,  the de Broglie duality relation  is a sufficient tool
to understand its nature and {\it the associated tachyonic matter wave
  is a reality not only to be reckoned with but also to be made use of
  in  understanding issues  where  we have  failed  with the  standard
  approach}.\\

When a wave packet is artificially constructed to force finitude and a
localised existence upon this quantum object, it inevitably spreads to
regain  its  infinitude  and  pervasiveness,  for  the  harmonic  time
dependence is  at the back of  both, the pervasiveness  (when space is
not  involved) and  the spreading  of  the wave  packet(when there  is
motion in space). \\

This  conclusion can  also be  inferred  from the  phenomenon of  {\it
  zitterbewegung} in relativistic quantum mechanics, where the attempt
to localize a particle beyond  a certain minimal limit invites zittery
oscillations making  the particle highly  unstable. So, it  seems that
{\it  the  concept of  a  localized particle  is  an  artifact of  our
  classical  outlook and  is not  in keeping  with the  nature  of the
  reality as  such}. As a result,  wherever we have  tried to forcibly
impose finitude  anywhere in quantum theory in  any manner whatsoever,
we  have  run into  insurmountable  conceptual difficulties  including
those connected with renormalization.\\

We shall look  at the issue of spreading a  little more closely taking
the relativistic mass energy formula in the following Section.\\

\noindent
{\bf 3. Group  velocity and the inevitable spreading  of wave packets}
\\

In  standard  quantum  theory, the  way  out  of  the problem  of  the
tachyonic phase velocity of the de Broglie waves is to superpose these
very waves with slightly differing  wavelengths so as to get some kind
of an  {\it average}  velocity by looking  at the stationarity  of the
phase around the central value $\vec{p}_{0}$ with respect to change in
momentum within  a small  range of values  $\Delta{p} \leq  |\vec{p} -
\vec{p_0}|$.  Such a wave packet can be written as:

\begin{equation}
\psi(\vec{r},t)=\int    \frac{d^3   p}   {(2\pi\hbar)^3}\phi(\vec{p})~
exp\left[\frac{i}{\hbar}(\vec{p}.\vec{r} - E(\vec{p})t)\right]
\label{pack}
\end{equation} 
where $\phi (\vec{p})$  is the weight function for  the momentum space
distribution.  Suppose  $\phi(\vec{p})$ has a small  range of non-zero
values in a region $\Delta p\leq|\vec{p}-\vec{p}_{0}|$ about a maximum
at $\vec{p}_{0}$.   The group velocity is calculated  by demanding the
stationarity                of                the               phase:
$\vec{\nabla}_{\vec{p}}\{{\vec{p}.\vec{r}-E(\vec{p})t}\}|_{\vec{p}_{0}}=0$,
which      yields      $\vec{r}(t)=v_{0}t$     with      $\vec{v}_{0}=
\vec{\nabla}_{\vec{p}}E(\vec{p})|_{\vec{p}_{0}}$.     This    position
$\vec{r}(t)=\vec{v}_{0}t$ corresponds to the maximum of $\psi(\vec{r},
t)$. It describes  the motion of the approximate  center of the packet
which we take  to represent the classical motion  of the particle with
velocity $\vec{v_0}$.  For an explicit evaluation of  the spreading in
time  of   the  packet,  we  take  the   relativistic  energy  formula
$E(\vec{p})=\sqrt{p^2c^2+m^2c^4}=\gamma  m c^2$  and expand  the phase
about $\vec{p}_{0}$ in a Tailor series to obtain:
\begin{eqnarray}
\vec{p}.\vec{r}-E(\vec{p})t=\vec{p}_{0}.\vec{r}-E(\vec{p}_0)t         +
(\vec{r}-\vec{\nabla}_{\vec{p}}E(\vec{p})t)|_{\vec{p}_{0}}.(\vec{p}-\vec{p}_0)
~~~~~~~~~~~~~~~~~~              \nonumber             \\             +
\frac{1}{2}\sum_{i,j}\left(-\frac{{\partial^2}       E}      {\partial
  p_i\partial              p_j}t\right)             \Bigg|_{\vec{p_0}}
(p_i-p_{0i})(p_j-p_{0j})+...      ...       ~~~~~~~~~~~      \nonumber
\\ =\vec{p}_0.\vec{r}-E(\vec{p}_{0})t+(\vec{r}-\vec{r}(t)).(\vec{p}-\vec{p}_0)
~~~~~~~~~~~~~~~~~~~~~~~~~~           \nonumber           \\          -
\frac{tc^2}{2E_0}\sum_{i,j}\left(\delta_{ij}-c^2\frac{p_{0i}p_{0j}}{E_{0}^2}\right)
\Delta{p_i}\Delta{p_j}+ ..... ~~~~~~~~~~~~~~~~~~~ \nonumber
\end{eqnarray}
 
where
\begin{eqnarray}
E_0=E(\vec{p}_0)=\sqrt{p_0^2c^2+m^2c^4}=\gamma_0mc^2
=\frac{mc^2}{\sqrt{1-\frac{v^2}{c^2}}}                               ~,
~~~~~~~~~~~~~~~~~~~~~~~\nonumber      \\      p_i-p_{0i}=\Delta{p_i},~
p_j-p_{0j}=\Delta{p_j};~                                  \vec{v}_g(t)=
\vec{\nabla}_{\vec{p}}E(\vec{p})|_{\vec{p}_{0}}       ~{\rm      and}~
\vec{r}(t)= \vec{r}_0(t)+\vec{v}_gt. ~~~~~ \nonumber
\end{eqnarray}
Substituting back in eq.(\ref{pack}) above we have,
\begin{eqnarray}
\psi(\vec{r},t)=exp\left[\frac{i}{\hbar}(\vec{p}_0.\vec{r}-E(\vec{p}_0)t)\right]
\int\frac{d^3p}{2\pi\hbar^3}\phi(\vec{p})
~exp~[\frac{i}{\hbar}\{(\vec{r}-        \vec{r}(t)).(\vec{p}-\vec{p}_0)
  ~~~~~~~~~ \nonumber\\ - \frac{tc^2}{2E_0}\sum_{i,j}\left(\delta_{ij}
  - c^2\frac{p_{0i}p_{0j}}{E_{0}^2}  \right)  \Delta{p_i}\Delta{p_j} +
  .....\}]
\end{eqnarray}

Without any loss  of generality, we now reorient our  axes so that the
motion of the packet is along the +ve x-direction. On writing \\
\hspace*{1in}$p=p_x,~  p_{0x}=p_0~$and $~E(p_{0x})=E(p_0)=E_0$,  \\ we
have
\begin{eqnarray}
\psi(x,t)=exp\left[\frac{i}{\hbar}(p_0x-E_0t)\right]\int\frac{dp}{2\pi\hbar}
\phi(p)~exp~[\frac{i}{\hbar}\{(x-x(t))(p-p_0)~~~~~~~~~~~~~~~~~~~~
  \nonumber \\ - \frac{tc^2}{2E_0} \left(1-\frac{c^2p_0}{E_0^2}\right)
  (p-p_0)^2+                                          .....\}]\nonumber
\\ =exp\left[\frac{i}{\hbar}(p_0x-E_0t)\right]\int\frac{dp}{2\pi\hbar}\phi(p)
~exp\left[\frac{i}{\hbar}(x-v_gt)(p-p_0)-\frac{it}{2\mu\hbar}(p-p_0)^2+
  .....\right]
\end{eqnarray}
where,
\begin{eqnarray}
\mu^{-1}=\frac{c^2}{E_0}\left(1-\frac          {c^2p_0^2}{E_0^2}\right)
\nonumber
\end{eqnarray}
 For definiteness, we now choose a Gaussian form for $\phi(p)$:
\begin{eqnarray}
\phi(p)=A~exp\left[-\frac{d^2}{\hbar^2}(p-p_0)^2\right]
\end{eqnarray}
Then the wave function on neglecting third order and higher becomes
\begin{eqnarray}
\psi(x,t)=exp\left[\frac{i}{\hbar}(p_0x-E_0t)\right]\frac{A}{2\pi\hbar}
\int dp~ exp\left[\frac{i}{\hbar}(x-v_gt)(p-p_0)-a(p-p_0)^2\right]
\end{eqnarray}
where,
\begin{eqnarray}
 a=\frac           {d^2}{\hbar^2}+\frac{it}{2\mu\hbar},~v_g          =
 \frac{\partial{E}}{\partial{p}}\Bigg|_{p_0} = \frac{p_0c^2}{E_0}.
\end{eqnarray}

To  further  simplify, we  put  $p_0x-E_0t=\phi_0$ and  $x-v_gt=\delta
x(t)=\delta x$ to obtain
\begin{eqnarray}
\psi(x,t)=exp\left(\frac{i}{\hbar}\phi_0\right)\frac{A}{2\pi\hbar}
\int                dp                ~exp\left[\frac{i}{\hbar}(\delta
  x(p-p_0)-a(p-p_0)^2\right]~~~~~~~~~~~~~~~~~                 \nonumber
\\  =exp\left(\frac{i}{\hbar}\phi_0 - ap_0^2  -\frac{i}{\hbar}\delta x
p_0\right)\frac{A}                 {2\pi\hbar}\int                 dp~
exp\left[{-ap^2+2a\left(p_0+\frac{i\delta x}{2a\hbar}\right)p}\right]
\end{eqnarray}

which, on evaluation yields
\begin{eqnarray}
\psi(x,t)=\sqrt{\frac{\pi}{a}}\frac{A}{2\pi\hbar}~exp\left[{\frac{i}{\hbar}}
  \phi_0-a\left(\frac{\delta {x}}{2a\hbar}\right)^2\right]~~~~~~~~~~~~
\nonumber\\ =\sqrt{\frac{\pi}{a}}\frac{A}{2\pi\hbar}exp\left[{\frac{i}{\hbar}}
  (p_0  x-E_0  t)-a\left(\frac{\delta  {x}}{{2a\hbar}}\right)^2\right]
\nonumber
\end{eqnarray}
The density is given by
\begin{eqnarray}
 |\psi(x,t)|^2=\left(\frac{A}{2   \pi   \hbar}\right)^2  \frac{\pi}{a}
 ~exp\left[2Re \left\{\frac{-(\delta x)^2}{4a\hbar^2}\right\}\right] =
 \left(\frac{A}{2\pi\hbar}\right)^2\frac{\pi}{a}
 ~exp\left[2Re\left\{\frac{-(\delta
     x)^2}{2d^2(1+\alpha^2)}\right\}\right]
\end{eqnarray} 
where, $\alpha=\frac{t\hbar}{2\mu  d^2}$ and the  normalisation factor
is $A=(8\pi d^2)^{1/4}$.  Thus we get the normalised density:
 \begin{equation}
|\psi(x,t)|^2=\frac{1}{d\sqrt{2\pi(1+\alpha^2)}}       ~exp\left[-\frac
  {(x-v_gt)^2}{2d^2(1+\alpha^2)}\right]\\
\end{equation}
 which  is a  Gaussian in  space whose  maximum moves  with  the group
 velocity
\begin{equation}
 v_g=\frac{\partial E}{\partial p}\Bigg|_{p_0}=\frac{p_0c^2}{E_0}
\end{equation}
Since the quantity
\begin{equation} 
\alpha=\frac{t\hbar}{2\mu        d^2}=\frac{t\hbar       c^2}{2d^2E_0}
\left(1-\frac{v_g^2}{c^2}\right)
\end{equation}
 increases linearly with time, the wave packet spreads.\\

 To compare with the corresponding nonrelativistic(NR) result obtained
 by  taking only  the kinetic  energy  in the  hamiltonian we  rewrite
 $\alpha$ by using $ E_0=\gamma mc^2$ as
\begin{eqnarray} 
\alpha=\frac{t\hbar}{2md^2}\left(1-\frac      {v_g^2}{c^2}\right)^{3/2}
=\alpha_{_{NR}} \left(1-\frac {v_g^2}{c^2}\right)^{3/2},
\end{eqnarray}
 where     $\gamma=\left(1-\frac     {v_g^2}{c^2}\right)^{1/2}$    and
 $\alpha_{NR}=\frac{t\hbar}{2md^2}$  is the  corresponding  factor for
 the nonrelativistic packet.  We  see that for faster moving particles
 with   $\beta=\frac{v}{c}    >>\frac{v_{_{NR}}}{c}$,   the   position
 uncertainty defined by
\begin{equation}
<\Delta x> = \sqrt{<(x-<x>)^2>}=d\sqrt{1+\alpha^2}
\end{equation}
grows  at  a lesser  and  lesser rate  compared  to  the low  velocity
nonrelativistic regime.   Nevertheless, spreading  is there and  it is
clear that the spreading is  an inevitable consequence of the harmonic
nature of the  time dependence of plane waves,  whatever be our choice
of the momentum distribution $\phi(p)$.\\

In the above  consideration, we have not kept  beyond the second order
terms in the  tailor series expansion of the phase,  but, we can still
get   the   ultrarelativistic    behaviour   by   taking   the   limit
$v_g\rightarrow c$.   We see that  for photons there is  no spreading,
which is  expected because of the  equality of the  phase velocity and
the group velocity for them.\\

\noindent
{\bf 4. The nonrelativistic group velocity}\\

Textbooks     on     nonrelativistic     quantum    mechanics     (see
e. g.  \cite{schwable}) describe the relation between  phase and group
velocities in the following manner:\\

The  energy  of  the  free   particle  is  entirely  kinetic  i.e.   $
E=\frac{1}{2}m v^2$.  The phase velocity  of the wave is $v_w =\omega/
k= E/p=\frac{p}{2m}=v/2$  while the group velocity of  the wave packet
is $v_g=\frac{d\omega}{dk}  =\frac{dE}{dp}=p/m=v= $ particle velocity.
This implies that the group velocity is twice the phase velocity. \\

However, the fact  that the general relation $v=pc^2/E$  holds for all
velocities  and that  the most  general expression  for the  energy is
given  by  the  relativistic  formula  $E^2=p^2c^2+m^2c^4$,  we  have,
$EdE=c^2pdp$ which gives the group velocity to be
\begin{equation}
v_g=\frac{dE}{dp}=\frac{pc^2}{E}=v
\end{equation}
and the phase velocity as
\begin{equation}
v_w=\frac{E}{p}=\frac{c^2}{v}
\end{equation}\\

Of  course, from  these  two  equations we  get  the general  relation
$v.v_w=c^2$  derived earlier  in Section  -  2. But,  if $v_w=v/2$  as
derived above, then there is a contradiction.\\

Our  emphasis  is  that  the  matter waves  are  always  tachyonic  in
character which means that we should have in all situations $v_w > v =
v_g$.  It is not  that this  contradiction has  gone unnoticed  in the
literature, although  textbook authors have  mostly gone by  the above
derivation and concluded unanimously that in the non-relativistic case
the phase  velocity is half  the group/particle velocity. In  a recent
article \cite{zumul},  it has  almost been figured  out but  not quite
remedied in a proper manner. \\

To  resolve  it,  we  begin  with the  {\it  correct}  nonrelativistic
expression for the energy of the free particle which {\it must include
  the rest energy}~:
\begin{equation}
E=mc^2+\frac{p^2}{2m}
\end{equation}
which gives for the phase velocity
\begin{equation}
v_w=\frac{E}{p}=\frac{p}{2m}+\frac{mc^2}{p}\\ =\frac{v}{2}+\frac{c^2}{\gamma
  v}\\ =\frac{v}{2}+\frac{c^2}{v}\left(1-\frac{v^2}{c^2}\right)^{1/2}
\end{equation}\\
 
When the term within brackets  is expanded and terms upto second order
are retained, it yields the correct relation. Further, we see that the
nonrelativistic energy  expression inclusive  of the rest  energy does
not alter  the result $v_g=v$  obtained above, as expected  because of
the constancy of the rest energy.\\

The  above analysis, in  addition to  removing a  common misconception
regarding the nonrelativistic limit  of the relation $v.v_w=c^2$, also
gives  us  a hint  that  there  is a  much  deeper  connection of  the
relativity theory  with the quantum theory \cite{jorgen},  at least as
far as de Broglie's approach is concerned. \\

In the following Section, we  further investigate this connection in a
Compton effect-like  situation and at the  same time bring  out from a
quite   different  perspective  the   robustness  of   the  expression
$v.v_w=c^2$ which leads  us to a reinterpretation of  our common sense
notion of ``{\it the energy carried by a particle}".\\

\noindent
{\bf 5. The robustness of ${\bf v. v_w = c^2}$ : a fresh approach }\\

In a  Compton Scattering experiment,  let an energy $\hbar  \omega$ be
absorbed by an electron(rest mass m) at rest. We note that this energy
is not the  whole energy of the incident photon but  is only that part
of the incident photon energy that  is taken up by the electron in the
process. As  a result  the electron would  move with a  kinetic energy
$K=\gamma mc^2-mc^2=(\gamma-1)mc^2$.  This  kinetic energy is entirely
due to the energy $\hbar \omega$ absorbed by it.  Thus $\hbar \omega =
(\gamma-1)mc^2$.  The total  energy  is then  $  E=\hbar \omega  +mc^2
=\hbar(\omega+\omega_0)$,  where   $\hbar\omega_0=mc^2$  is  the  rest
energy.  Thus,
\begin{equation}
E=\sqrt{p^2c^2+m^2c^4}=\hbar\omega+mc^2
\end{equation}
 whence,
\begin{equation}
p=\frac{\hbar\omega}{c}\sqrt{1+\frac{2mc^2}{\hbar\omega}} =\frac{\hbar
  \omega}{c}\sqrt{1+\frac{\omega_z}{\omega}}
\end{equation} 

where, $\omega_z=2mc^2/\hbar$ is the {\it zitterbewegung} frequency of
the particle.  Using  this and the expression for  the total energy of
the particle we deduce for the phase velocity ,
\begin{equation}
v_w=\frac{E}{p}
=c\frac{1+\frac{\omega_z}{2\omega}}{\sqrt{1+\frac{\omega_z}{\omega}}}
=\frac{c^2}{v}\geq c
\end{equation}

where,  the equality holds  only for  massless particles.  Writing the
Compton wavelength as $\lambda_c=\frac{h}{mc}$ and the incident photon
wavelength as $\lambda_\gamma=2\pi c/\omega$, we can explicitly verify
that the  de Broglie wavelength ${\lambda_{dB}}=h/p$ is  related to it
by:
\begin{equation} 
\lambda_{dB}=\lambda_{\gamma}\left(1+\frac{\omega_z}{\omega}\right)^{-1/2}
=\lambda_{\gamma}\sqrt{1+\frac{2\lambda_{\gamma}}{\lambda_c}}
\end{equation}
 which is  always less than  the absorbed photon wavelength  while the
 frequency     of     the     matter     wave     $\nu_{dB}=\nu_\gamma
 (1+\omega_z/2\omega)$  is always  greater than  the frequency  of the
 absorbed photon  so that $  v_w=\lambda_{dB}~ \nu_{dB}=c^2 /v\geq{c}$
 holds always i. e. the matter wave is tachyonic.\\

Thus,  we have  derived  the relation  from  the point  of  view of  a
particle  absorbing  an amount  of  energy  $\hbar\omega$  and in  the
process undergoing a change of state from rest to motion with velocity
$v$.\\

This discussion gives us a  novel understanding of the motional energy
or kinetic energy.   Usually, we express ourselves by  saying that the
particle  is possessing  or  carrying kinetic  energy.   But, here  we
clearly  see  that  we can  very  well  say  from a  more  fundamental
viewpoint that it  is not that the particle  is possessing or carrying
the (photon)kinetic energy, rather, it  is the photon (which is always
in motion with speed c)that is carrying or possessing the particle and
that has  become burdened by  the particle of  mass m. As a  result of
which, it is able to carry the  latter only with a speed less than its
original speed  c which we  identify as the bradyonic  particle.  When
this  occurs,  we have  the  pervasive  tachyon  corresponding to  the
particle at  rest going over  to the tachyonic  matter wave or  the de
Broglie wave train along the line of motion.\\

Accordingly, we  can define a single-particle  refractive index $\eta$
as follows:
\begin{equation} 
\eta= \frac{c}{v_w}=\sqrt{1-\frac{\omega_0^2}{\omega^{{\prime}^{2}}}}
\end{equation} 
where, $\omega^\prime=E/\hbar$ with  $E=\hbar\omega+mc^2$ as the total
energy. We see further that  this refractive index is always less than
one, which  means that  the matter wave  moves with velocity  $ v_w>c$
when the photon is loaded with the particle of mass m.\\

Moreover, we  wish to point out  that the phase velocity  can never be
less than the  particle velocity and that the  several consistent ways
of deriving the  same reciprocal relation between the  two go to prove
without  doubt  that the  matter  waves  are  always tachyonic  for  a
bradyonic particle. We note also that the quantum object at rest being
an   all-pervading  existence,   has  actually   no   classical  point
particle-like  character because it  is a  tachyonic wave  of infinite
wavelength before measurement.  The manifestation of this may be taken
to be in the form of  the the transverse longitudinal fields (like the
gravitational or electrostatic) \cite{feyn,walker} associated with the
particle, which have been  found to propagate with infinite velocities
tallying   fully   well    with   the   ``{\it   pervasive   existence
  interpretation}''  of the quantum  object at  rest proposed  in this
article. \\

For example, for an electron  at rest, its electric field extends upto
infinity and no  matter where in space another  charge is situated, it
interacts with the latter- in  principle at least- though the strength
decreases  as   $1/(distance)^2$.  Even  this   inverse  square  field
dependence may very well be a fact only for interacting sources, since
we've no way of experimentally  determining the field of a free source
i.e. without  subjecting it to  an interaction.  The assigning  of the
$1/(distance)^2~$  field dependence to  the free  particle is  only an
extrapolation   of    the   interacting   field-dependence    to   the
non-interacting case.   Here, in this clssical  extrapolation lies the
crux  of the  matter, the  root of  all the  problems faced  by  us in
tackling  self-interaction   etc.  which   forced  us  to   resort  to
renormalization in  QFT. In fact, it  has already been  proposed in the
literature\cite{chuby}  that  the   longitudinal  and  the  transverse
aspects of  the electromagnetic field may  be thought of  as dual, and
hence complementary, to  each other and together they  are to be taken
to make up the full Reality of the electromagnetic field. \\

Thus  we may  say  that the  concept  of a  localised  particle is  an
artefact  of our classical  imagination and  is deduced  from everyday
experience  with  localized  macroscopic  objects  and  therefore,  is
divorced from the microscopic reality of the quantum world.\\

\noindent
{\bf 6. Bradyon-Tachyon Duality/Complementarity}\\

Basing on the above discussion we extend the complementarity principle
of Bohr  to say that  ``the bradyonic and  the tachyonic aspects  of a
quantum  object are  complementary to  each other''.  They may  not be
simultaneously  observable in  a  single experiment  and  we may  need
appropriately designed  experiments separately for  the observation of
any one aspect. Locality and non-locality, causality and teleology may
be similarly thought  of as making up a  fuller reality, more complete
than hitherto accepted  in Physics.  We may be  just passing through a
similar phase  of reconciling dualities via  a complementary principle
as happened in the early days of Quantum theory a century ago.\\

Thus, we  propose that Quantum  Reality has a  complementary tachyonic
aspect associated with every bradyon as follows:

$\bullet $ Bradyon at rest $\Leftrightarrow$ pervasive tachyon.

$\bullet$  Bradyon in  motion $\Leftrightarrow$  tachyonic  wave train
along the line of motion.

Therefore, we rewrite the Einstein-de Broglie relation in the form
\begin{equation}
v_b.v_t=c^2
\end{equation}
where, $v_b$  and $v_t$ are  the velocities of the  bradyonic particle
and its associated tachyonic matter wave. This relation is to be taken
as the  starting point for the  investigation of the  possible role of
tachyons in explaining quantum entanglement and other related nonlocal
phenomena. The reason why we do not observe tachyons is because of our
presumption that they are  localised objects like bradyons, which they
are not.  They being fundamentally  pervasive in character,  we cannot
detect them  experimentally in the traditional sense,  but the duality
relation above  may be  taken to  be give us  indirect proof  of their
existence.\\

We see  that if we  accept the viewpoint  advocated in this  paper, we
immediately grant a reality to  the tachyons which is long overdue. In
fact,  we have all  along been  working with  them since  de Broglie's
original work in the form of  matter waves ! Though the above relation
looks  like  a  restatement  of  the  wave-particle  duality  and  the
conjugate-variable   complementarity  in   old  quantum   theory,  the
conceptual shift  in the paradigm is  a huge one as  discussed in this
work.  In  fact,  with  the  interpretation  proposed  here,  the  old
bradyon-tachyon  complementarity  which was  earlier  required of  the
relativity  theory for  its  completeness in  the  scheme of  extended
relativity proposed  by Recami and  others \cite{recam} becomes  now a
point  of conformity  with  quantum  theory, in  the  sense that  both
theories have the tachyons included as essential constituents. \\

We    note    that   while    in    all    the   quantum    mechanical
duality/complementarity relations it is Planck's constant $h$ which is
the fundamental constant linking  the two aspects (e.g.$~\lambda.p = h
$),  here $c^2$  is the  reciprocality constant  connecting  $v_b$ and
$v_t$. Interestingly, the  equation (27) can also be  cast in terms of
the dimensionless boost parameters:
\begin{equation}
\beta_b.\beta_t=1
\end{equation} \\ 
The reciprocality is such that in  order not to be at loggerheads with
Relativity,  we have  to  accept  that {\it  tachyons,  by their  very
  nature, are pervasive and that there can be no point-like tachyons}.
The  notion of localised  point-like particles  is applicable  only to
bradyons.  Thus,  rather than trying to rule  out quantum non-locality
due to its conflict with relativity,  we now rest on a solid ground of
unification on the basis  of the bradyon-tachyon complementarity where
both the theories have non-locality as a common characteristic inbuilt
into their structure through the tachyons.\\

\noindent 
{\bf 7. Discussion and conclusion} \\

We've shown that the pre-measurement state of a free quantum object at
rest  is not the  same as  the post-measurement  state of  a localized
particle-like   existence,   but  is   a   pervasive  existence.   The
localization that we are  familiar with from our classical observation
of particulate existences  can be interpreted to be  the result of the
interaction of  the measuring apparatus  with the free  quantum object
which brings  about the collapse  of its pervasiveness to  a pointlike
existence   by  the   very  design   of  our   experiments   or  other
interactions. We've shown that the wave packet description is not free
from  the  problem of  spreading  even  when  the relativistic  energy
formula  is employed. The  wave packet  description is  shown to  be a
post-measurement description  incorporating our classical common-sense
notion  of a localised  pointlike particle  and thus  is not  the true
description  of  the  free  quantum  object  as  it  is  i.e.   before
measurement.  The relation between  the group and the phase velocities
is  derived   using  different   approaches  and  its   robustness  is
established.\\

A  long-standing  conceptual error  in  textbooks  is  pointed out  in
connection  with the nonrelativistic  limit of  this relation.  In our
approach  we  have  shown  that  including  the  rest  energy  in  the
expression  for total energy  remedies the  situation and  proves once
again the  robustness of the relation  beyond any element  of doubt as
well as its universal validity. \\

Basing on  this relation  we have proposed  a reinterpretation  of the
notion of the kinetic energy {\it carried} by a particle. We have also
proposed to extend Bohr's complementarity principle to bradyon-tachyon
complementarity thereby  giving the  tachyons their rightful  place in
the scheme of  quantum mechanics.  The familiar point  particle is but
the `tip  of the iceberg' of pervasive  tachyonic existence associated
with it.\\

It is worth noting that the nonlocal EPR-like quantum correlations and
entanglement  effects  become easy  to  understand,  once a  pervasive
existence is granted to the  quantum objects at rest and an associated
tachyonic matter wave is granted to the moving quantum particles.  The
basic objection to the existence of tachyonic waves that if they exist
they would carry  signals faster than light can be  met by the equally
basic  fact   that  a  tachyonic   matter  wave  being   extended  and
encompassing in nature does not  need to carry any information between
two points as it simultaneously touches both ends! \\

The  present  work  vindicates  the  efforts in  some  of  the  recent
work\cite{recent} where  the authors have also argued  for the reality
of  the de  Broglie waves.  However,  there remain  many issues  still
unsolved  regarding  the  exact  description  of  the  interaction  of
tachyons and bradyons and  amongst tachyons themselves; the connection
of  the  tachyonic  matter  waves with  the  antiparticles,  spacelike
measurements,  and  finally,  with  consciousness  which  need  to  be
explored in future work.\\

\noindent
  {\bf Acknowledgments}\\

The first author wishes  to thankfully acknowledge the invaluable help
rendered  by  Dr. J.   Kamila  and by  Sri  Artabandhu  Mishra in  the
preparation of the  paper and also to thank  the Institute of Physics,
Bhubaneswar for providing the computer and library facilities.


\begin{thebibliography}{99} 
\bibitem{recami}Erasmo Recami, Flavio  Fontana and Roberto Garavaglia,
  International Journal of Modern Physics A15 (2000) 2793-2812

\bibitem{asoke} Ashoke Sen, Int.J.Mod.Phys. A20 (2005) 5513-5656

\bibitem{infl} D. A. Steer and F.Vernizzi, Phys.Rev.D70:043527,2004
 
\bibitem{padman} P. C. W. Davies, Int.J.Theor.Phys. 43 (2004) 141-149
\bibitem{reca} V.  S. Olkhovsky, E.  Recami and J.  Jakiel,phys. rep.,
  398(2004),133-178.
\bibitem{deBrog}  L.  de  Broglie,  “Waves  and  quanta”,  Nature  112
  (19230-543.
\bibitem{expt}   A.M.Steinberg,  P.   G.  Kwiat   and  R.   Y.  Chiao,
  Phys.  Rev.  Lett.  71 (1993)  708;  R.  Y.  Chiao, P.G.  Kwiat  and
  A.  M.  Steinberg:  Scient.   Am.  269  (1993),  issue  no.2,  p.38;
  Ph. Balcou and L. Dutriaux: Phys. Rev. Lett. 78 (1997) 851; V. Laude
  and P. Tournois: J. Opt. Soc. Am. B16 (1999) 194.
\bibitem{bohm} D. Bohm, Phys. Rev. 85 (1952) 166.
\bibitem{bohm1} R. Feynman, Rev. Mod. Phys. 20 (1948) 367.
\bibitem{qm} H. Everett, Reviews of Modern Physics, 29 (1957) 454.
\bibitem{debro} L. de Broglie, Ph. D. thesis, Univ. Of Paris, (1924).
\bibitem{schwable}   F.   Schwable,   ``Quantum  Mechanics",   Springer
  Int. Students' Edn., Narosa Pub. House, (1995),pp-15.
\bibitem{zumul} Yusuf Z. Umul, arxiv:0712.0967
\bibitem{jorgen}    T.    P.    Jorgensen,    Int.J.Theor.Phys.    37,
  2763-2766,(1993)
\bibitem{feyn}  R. P. Feynman  in ``The  Feynman Lectures  in Physics",
  Vol.2, Addison Wesley, Ch.21,(1989)
\bibitem{walker} W. D. Walker and J. Dual, arxiv:gr-qc/9706082.
\bibitem{chuby}A.   Chubykalo,   A.   espinoza,  alvarado-Flores   and
  Alejandro   Guitierrez,Found.   Phys.   Lett.,19,1,   Pp.37-49   and
  references therein.
\bibitem{recam}   E.  Recami,  Rivista   Nuovo  Cimento   9,  1(1986);
  O.  M.  P.  Bilaniuk, V.  K.  Deshpande,  and  E. C.  G.  Sudarshan,
  Am. Journ. Phys.  30, 718 (1962); G. Feinberg,  Phys. Rev. 159, 1089
  (1967)
\bibitem{recent} A. p. Kirilyuk, arXiv: quant-ph/991107; J. X. 
  Zheng-Johansson  and P-I. Johansson,  Prog. in Phys. V.4, 32, 2006.



\end{thebibliography}
\end{document}